\documentclass[article,shortnames,nojss]{jss}

\usepackage{orcidlink,thumbpdf,lmodern}

\usepackage{framed}

\newcommand{\class}[1]{`\code{#1}'}
\newcommand{\fct}[1]{\code{#1()}}

\usepackage{amsmath,amsfonts,amssymb}
\DeclareMathOperator{\cov}{cov}

\newcommand{\ie}{I\!E}

\newcommand{\mb}[1]{\mbox{\boldmath \ensuremath{#1}}}

\author{Yan-Feng Wu~\orcidlink{0000-0002-7105-1070}\\Fudan University
   \And Jian-Qiang Hu~\orcidlink{0000-0003-2989-8048}\\Fudan University}
\Plainauthor{Yan-Feng Wu, Jian-Qiang Hu}

\title{ajdmom: A \proglang{Python} Package for Deriving Moment Formulae of Affine Jump Diffusion Processes}
\Plaintitle{ajdmom: A Python Package for Deriving Moment Formulae of Affine Jump Diffusion Processes}
\Shorttitle{\pkg{ajdmom}: Derivation of Moment Formulae for Affine Jump Diffusion in \proglang{Python}}

\Abstract{
  We introduce \pkg{ajdmom}, a \proglang{Python} package designed for automatically deriving moment formulae for the well-established affine jump diffusion processes with state-independent jump intensities. \pkg{ajdmom} can produce explicit closed-form expressions for conditional and unconditional moments of any order, significantly enhancing the usability of these models. Additionally, \pkg{ajdmom} can compute partial derivatives of these moments with respect to the model parameters, offering a valuable tool for sensitivity analysis. The package's modular architecture makes it easy for adaptation and extension by researchers. \pkg{ajdmom} is open-source and readily available for installation from GitHub or the \proglang{Python} package index (PyPI).
}

\Plainkeywords{Affine jump diffusion, moment derivation, Heston model, stochastic volatility, Python}

\Address{
  Yan-Feng Wu, Jian-Qiang Hu\\
  Department of Management Science\\
  Fudan University\\
  No. 670, Guoshun Rd.\\
  200433 Shanghai, China\\
  E-mail: \email{wuyf@fudan.edu.cn}, \email{hujq@fudan.edu.cn}\\
  URL: \url{https://github.com/xmlongan}
}

\begin{document}



\section[Introduction]{Introduction} \label{sec:intro}
Affine jump diffusion (AJD) processes are distinguished by their aptitude for capturing the intricate dynamics of various stochastic systems while preserving a measure of analytical tractability. These processes are characterized by affine dependencies of the drift, the instantaneous covariance matrix, and the jump intensity on state vectors, as described by \citet{duffie2000transform}. There are several well-known classes of AJD processes, including the Ornstein-Uhlenbeck (OU) process and the square-root diffusion (SRD) process, the latter of which was introduced by \cite{cox1985theory} for modelling the term structure of interest rates. The SRD model has  also been integrated into the stochastic volatility (SV) setting, notably in the Heston SV model for option pricing as formulated by \citet{heston1993closed}. This model has been further extended to incorporate jumps, addressing fat tails and volatility clustering through SV with jumps in returns (SVJ) and SV with contemporaneous jumps in returns and volatility (SVCJ), as investigated by \citet{bates1996jumps}, \citet{duffie2000transform}, \citet{eraker2003impact}, and \citet{broadie2007model}. Although these models have predominantly been applied in the realm of financial asset pricing and econometrics, their applications may extend to other areas such as inventory management (\citealt{canyakmaz2019inventory}), queueing systems (\citealt{weinberg2007bayesian}; \citealt{giesecke2011exact}), electricity markets, neural activity and others as \citet{gorjao2023jumpdiff} pointed out.

Despite their versatility, AJD processes generally lack closed-form transition and marginal densities---even for basic cases such as the SRD and Heston SV models. Consequently, deriving their moments and covariances becomes a central challenge in their study, as these are crucial for modeling and parameter estimation. As shown by \citet{kyriakou2023unified}, the probability distributions of these models (including Heston SV, SVJ, and SVCJ) are uniquely determined by their moments under certain regularity conditions. Their work also demonstrates that moment-based simulation can yield significant computational savings compared to conventional methods relying on conditional characteristic function (CF) inversion. Furthermore, the method of moments has been widely adopted for parameter estimation in AJDs, as evidenced by \citet{wu2019moment,yang2021method,wu2022moment,wu2024method,wu2025method}.

Until very recently, closed-form moments and covariances are only available for simple AJD models like SRD. The Heston SV model, for example, one of the most popular SV models, has only a known analytical conditional CF \citep{heston1993closed}, though \citet{jiang2002estimation} developed a closed-form (unconditional) CF for a simplified version of the model. \cite{bollerslev2002estimating} derived conditional moments for the integrated state of the Heston SV model and its AJD extensions with jumps. For general affine diffusion and AJD models, conditional CFs have been established by \cite{duffie2000transform}, \cite{singleton2001estimation}, and \cite{chacko2003spectral}. For polynomial processes, a broader class of models,  \citet{cuchiero2012polynomial} show that all finite-order conditional moments are analytically tractable, up to a matrix exponential. For a detailed discussion of moments in related processes, see \citet{wu2025density}. The exact simulation of the Heston SV model and its AJD extensions via numerical inversion of conditional CFs is studied by \cite{broadie2006exact}. \citet{kyriakou2023unified} propose a moment-based simulation procedure with moments evaluated from conditional CFs. Simulation of relevant models based on various approximations can be found in \cite{choi2024simulation, giesecke2011exact, zhang2015affine,dassios2017efficient}.

In our recent work, we developed a recursive procedure that can in principle be used to derive moments and covariances for AJD models with state-independent jump intensities \citep{wu2025density}. This marks the first time that fully explicit closed-form moments and covariances can be derived for a broad range of AJD models. Although this procedure can easily produce lower-order moments and covariances, it becomes extremely complex as the order of moments and covariances grows. For instance, computing the fourth moment for a very simple AJD model requires approximately $12^4 = 20,736$ sub-steps, which makes manual calculation highly impractical. In addition, calculations for different AJD models, such as those having different jump distributions, are usually quite varied. Therefore, it would be very helpful to develop a software tool to calculate the moments and covariances based on our method automatically.  

Symbolic computation is typically employed for algebraic operations, differentiation, integration, and polynomial factorization, with support provided by programming languages and libraries such as \proglang{Mathematica} \citep{Mathematica}, \proglang{MATLAB} \citep{matlab2022}, and \proglang{Python}'s \citep{van2009python} package \pkg{SymPy}  \citep{sympy2017}. However, to the best of our knowledge, none of these currently offers functionalities for manipulating Itô processes, which are essential to apply our method to calculating the moments and covariances of AJD models. 

Currently, there are some packages offering varied functionalities for different types of jump-diffusion processes. One of the most commonly supported functionalities is the simulation of these processes, mainly applied to financial option pricing. The \proglang{MATLAB} \citep{matlab2022} package \pkg{PROJ\_Option\_Pricing\_Matlab} \citep{Kirkby2021PROJ} supports option pricing for diffusions, jump diffusions, and some general Lévy processes models. Meanwhile, the \proglang{C++} \citep{stroustrup2013c++} project \pkg{DerivativesPricing} \citep{Gosain2020} and the \proglang{R} \citep{R2024} package \pkg{Jdmbs} \citep{okada2020Jdmbs} have implemented option pricing algorithms only for the classical Black-Scholes \citep{black1973pricing} geometric Brownian motion model and some simple extensions with jumps. The package \pkg{JumpProcesses.jl} \citep{Zagatti2024} is available in \proglang{Julia} \citep{bezanson2017julia} for simulating point processes with time-varying intensities, and the \proglang{Julia} package \pkg{DifferentialEquations.jl} \citep{DifferentialEquations.jl-2017} supports numerical solving of general diffusion processes. \citet{rinn2016langevin} developed an \proglang{R} package for modeling Markov processes based on the Langevin approach. There are two \proglang{Python} packages implementing non-parametric modeling of jump diffusion processes through the Kramers-Moyal equation \citep{rydin2021arbitrary}: \pkg{jumpdiff} \citep{gorjao2023jumpdiff} and \pkg{kramersmoyal} \citep{gorjao2019}. Although non-parametric models are more general compared with parametric ones, the AJD processes considered in our package \pkg{ajdmom} are parametric models and are mainly applied to option pricing, in which the parameters usually have specific economic meanings, therefore favored by economists for modeling certain economic behaviors. In order to recover the equations of the jump diffusion processes, the package \pkg{jumpdiff} uses conditional moments from data to infer the Kramers-Moyal coefficients. In contrast to the Markov settings in \pkg{jumpdiff}, our method contributes to the community by deriving both conditional and unconditional moment formulae for the AJD processes under hidden Markov model (HMM) settings. For instance, in the settings of SV models, the volatility process is usually unobservable.

This paper introduces \pkg{ajdmom}, a \proglang{Python} package we designed and implemented for automating the derivation of the moment and covariance formulae of AJDs with state-independent jump intensities---the most commonly used class of AJD models. The \pkg{ajdmom} package is predicated on the insight that all moments and intermediate conditional moments can be represented as generalized polynomials, leading us to create a custom dictionary data structure, the \class{Poly} class. This class facilitates auto-derivation of the moments and covariances for the AJD models. The package offers several features, including its support for deriving moments, central moments, and covariances of any order, partial differentiation with respect to model parameters, and a modular design that easily extends to other AJD or even some non-affine jump diffusion models.

The remainder of this paper is organized as follows: Section 2 provides an overview of AJD models and the recursive procedure used in deriving the moments in the baseline model \textemdash Heston SV model. Section 3 discusses the design and code structure of the \pkg{ajdmom} package, in which the AJD extensions of the Heston SV model are included. Section 4 demonstrates the application of \pkg{ajdmom} to the Heston SV, SVJ and SVCJ models through experiments that confirm the method's accuracy and implementation. Finally, Section 5 concludes the paper.
Readers who are familiar with AJD models but do not wish to delve into the details of moment derivation can skip Sections~2 and 3 and proceed directly to Section~4 for more generalized information. Comprehensive documentation of the package, built with the \proglang{Python} tool \pkg{Sphinx} \citep{sphinx_2020_sphinx}, is hosted at \url{http://www.yyschools.com/ajdmom}, and the source code repository is available on GitHub at \url{https://github.com/xmlongan/ajdmom}.




\section{AJD models and recursive equation} \label{sec:models}
The general AJD process is an n-dimensional Markov process, denoted by 
$\mb{x}(t)$ with state space $D\subset \mathbb{R}^n$, and evolves according to the stochastic differential equation (SDE):
$$
    d \mb{x}(t) = \mb{\mu}(\mb{x}(t))dt + \mb{\sigma}(\mb{x}(t))d \mb{w}(t) + d\mb{z}(t),
$$
where 
\begin{itemize}
    \item $\mb{w}(t)$ is an n-dimensional standard Wiener process,
    \item $\mb{z}(t)$ is an inhomogeneous compound poisson process (CPP) with jump distribution $F_{\mb{j}}(\cdot)$ on $\mathbb{R}^n$ and intensity $\boldsymbol{\lambda}(\mb{x}(t)): D\rightarrow \mathbb{R}_{\geqslant 0}$.
\end{itemize}
The drift $\mb{\mu}(\cdot)$, instantaneous covariance matrix $\mb{\sigma}(\cdot)\mb{\sigma}(\cdot)^T$, and jump intensity $\boldsymbol{\lambda}(\cdot)$ all have affine dependence on $\mb{x}(t)$ \citep{duffie2000transform}, parameterized by coefficients $(\boldsymbol{K},\boldsymbol{H},\boldsymbol{l})$ as follows:
\begin{itemize}
    \item $\boldsymbol{\mu}(\boldsymbol{x}) = \boldsymbol{K}_0 + \boldsymbol{K}_1\boldsymbol{x}$, for  $\boldsymbol{K} = (\boldsymbol{K}_0,\boldsymbol{K}_1)\in \mathbb{R}^n\times \mathbb{R}^{n\times n}$.
    \item $(\boldsymbol{\sigma}(\boldsymbol{x})\boldsymbol{\sigma}(\boldsymbol{x})^T)_{ij} = (\boldsymbol{H}_0)_{ij} + (\boldsymbol{H}_1)_{ij}\cdot \boldsymbol{x}$, for $\boldsymbol{H}=(\boldsymbol{H}_0,\boldsymbol{H}_1)\in \mathbb{R}^{n\times n}\times \mathbb{R}^{n\times n\times n}$.
    \item $\boldsymbol{\lambda}(\boldsymbol{x}) = \boldsymbol{l}_0 + \boldsymbol{l}_1\cdot \boldsymbol{x}$, for $\boldsymbol{l}=(\boldsymbol{l}_0,\boldsymbol{l}_1)\in \mathbb{R}^n\times\mathbb{R}^{n\times n}$.
\end{itemize}

AJD models have been predominantly applied in financial asset valuation and econometric analysis. In this paper, we concentrate on the most typical class of AJDs---those with state-independent jump intensities (i.e., $\boldsymbol{\lambda}(\boldsymbol{x}) = \boldsymbol{l}_0$). This class includes widely used models such as the Heston SV, SVJ, and SVCJ. It also encompasses other variants, including two-factor SV, two-factor SV with jumps in returns, SV with jumps in volatility, SV with independent jumps in returns and volatility, square-root jump diffusion, superposition of square-root diffusion and so on.

Among these AJDs, SV models present the greatest theoretical and computational challenges: their volatility component is inherently unobservable, and the observable price process violates the Markov property. Consequently, deriving their transition and marginal densities has long been problematic. Recent work by \citet{wu2025density} overcome this issue through a recursive methodology that yields closed-form expressions for conditional and unconditional moments of arbitrary order. In the rest of this section, we present this recursive approach for the Heston SV model as a baseline case; extensions to other AJDs are provided in \citet{wu2025density}. This theoretical framework underpins the design of the Python package \pkg{ajdmom}, which automates the derivation of moment formulae.

The Heston SV model, foundational to the \pkg{ajdmom} package, is formalized by the following system of SDEs \citep{heston1993closed}:
\begin{align}
    dp(t) &= (\mu -v(t))dt + \sqrt{v(t)}dw^s(t),\label{eqn:log-price}\\
    dv(t) &= k(\theta - v(t))dt + \sigma_{v}\sqrt{v(t)}dw^v(t),\label{eqn:v-t}
\end{align}
where $p(t)$ denotes the logarithm of asset price at time $t$, $\mu$ is a constant return, $v(t)$ is the instantaneous variance of return at time $t$, $w^s(t) = \rho w^v(t) + \sqrt{1-\rho^2}w(t)$, and $w^v(t)$ and $w(t)$ are two independent Wiener processes. The process $v(t)$ is an SRD, akin to the Cox-Ingersoll-Ross (CIR) process \citep{cox1985theory}. Parameters $k>0$, $\theta>0$, $\sigma_v>0$, and initial variance $v(0)>0$, and it is required to ensure the positivity of the variance process $v(t)$ that $2k\theta>\sigma_v^2$. For notational simplicity, the following integrals are introduced:
\begin{align*}
  I_{s,t}\equiv \int_s^t\sqrt{v(u)}dw^v(u),\quad
  I_{s,t}^* \equiv \int_s^t\sqrt{v(u)}dw(u),\quad
  I\!E_{s,t} \equiv \int_s^te^{ku}\sqrt{v(u)}dw^v(u).
\end{align*}
Subsequently, $I_{n,t} \equiv I_{nh, t}$, $I_{n,t}^{*} \equiv I_{nh,t}^{*}$, $\ie_{n,t} \equiv \ie_{nh,t}$, and $I_n \equiv I_{(n-1)h,nh}$, $I_n^{*}\equiv I^{*}_{(n-1)h,nh}$ and $\ie_n \equiv \ie_{(n-1)h,nh}$.

The return over the $n$-th interval of length $h$ is defined as $y_n \equiv p(nh) - p((n-1)h).$
To bridge discrete and continuous analysis, $y_{n-1,t}$ is introduced as an intermediary variable that captures the process dynamics between any time $t$ and the preceding discretization mark $(n-1)h$, $y_{n-1,t} \equiv p(t) - p((n-1)h)$.
Upon normalizing $y_{n-1,t}$ by its conditional expected value, with condition on $v_{n-1}$ ($\equiv v((n-1)h)$), we define the centralized process $\bar{y}_{n-1,t}$:
$$
    \bar{y}_{n-1,t} \equiv y_{n-1,t} - \E[y_{n-1,t}|v_{n-1}]
$$
which can be decomposed into its constituent terms as:
$$
    \bar{y}_{n-1,t}
    = \frac{\sigma_v}{2k} e^{-kt} I\!E_{n-1,t}
     + \left(\rho - \frac{\sigma_v}{2k} \right)I_{n-1,t} + \sqrt{1-\rho^2}I_{n-1,t}^* - \beta_{n-1,t}\bar{v}_{n-1} ,
$$
where coefficient $\beta_{n-1,t} \equiv (1-e^{-k[t-(n-1)h]})/(2k)$ and normalized variance $\bar{v}_{n-1} \equiv v_{n-1} - \theta$.
The moments of $\bar{y}_{n}$ can be computed based on the moments of $\bar{y}_{n-1,t}$ at $t=nh$, thereby simplifying the analysis to discrete time points. We will consider hereafter computing the moments of $\bar{y}_{n-1,t}$.  

The $m$-th central moment of $y_{n-1,t}$ can be derived by relying on a set of underlying components:
\begin{align}
    \E[I\!E_{n-1,t}^{m_1}I_{n-1,t}^{m_2}I_{n-1,t}^{*m_{3}} \bar{v}_{n-1}^{m_4}],\label{eqn:comb-moment}
\end{align}
where $m_i\geqslant 0$ for $i=1,2,3,4$ and the summation $\sum_{i=1}^{4}m_i=m$. The component moment \eqref{eqn:comb-moment} can be computed via a two-step process:
$$
\E[\E[I\!E_{n-1,t}^{m_1}I_{n-1,t}^{m_2}I_{n-1,t}^{*m_{3}}|v_{n-1}] \bar{v}_{n-1}^{m_4}].
$$
i.e., first compute the inner conditional expectation, then follow with the unconditional expectation. 
\citet{wu2025density} construct a recursive equation for deriving the conditional product moment:
\begin{align}\label{eqn:recursive-3items}
    &\E[\ie_{n-1,t}^{m_1}I_{n-1,t}^{m_2}I_{n-1,t}^{*m_3}|v_{n-1}]\nonumber\\
    &= f(\E[\ie_{n-1,s}^{m_1-2}I_{n-1,s}^{m_2}I_{n-1,s}^{*m_3}|v_{n-1}], \E[\ie_{n-1,s}^{m_1-1}I_{n-1,s}^{m_2}I_{n-1,s}^{*m_3}|v_{n-1}],\nonumber\\
    &\quad\quad~\E[\ie_{n-1,s}^{m_1}I_{n-1,s}^{m_2-2}I_{n-1,s}^{*m_3}|v_{n-1}],\E[\ie_{n-1,s}^{m_1+1}I_{n-1,s}^{m_2-2}I_{n-1,s}^{*m_3}|v_{n-1}],\nonumber\\
    &\quad\quad~\E[\ie_{n-1,s}^{m_1-1}I_{n-1,s}^{m_2-1}I_{n-1,s}^{*m_3}|v_{n-1}],\E[\ie_{n-1,s}^{m_1}I_{n-1,s}^{m_2-1}I_{n-1,s}^{*m_3}|v_{n-1}],\nonumber\\
    &\quad\quad~\E[\ie_{n-1,s}^{m_1}I_{n-1,s}^{m_2}I_{n-1,s}^{*m_3-2}|v_{n-1}],\E[\ie_{n-1,s}^{m_1+1}I_{n-1,s}^{m_2}I_{n-1,s}^{*m_3-2}|v_{n-1}]),
\end{align}
where function $f$ is defined in the Appendix. The final formulae of the conditional product moments can be expressed as polynomials in $v_{n-1}$. This implies that the moment described in Equation~\eqref{eqn:comb-moment} can be represented in terms of the moments of $v_{n-1}$. 
Meanwhile, the $m$-th moment of $v_{n-1}$ is given by
\begin{align}
    \E[v^m_{n-1}] = \prod_{j=0}^{m-1} \left(\theta + \frac{j\sigma_v^2}{2k} \right), \quad m=1,2,\dots,\label{eqn:v-moment}
\end{align}
under the strict stationarity assumption of $v(t)$ \citep{overbeck1997}.
This solution emerges because the stationary process $v(t)$ follows a gamma distribution with mean $\theta$ and variance $\theta \sigma_v^2/(2k)$ \citep{cox1985theory}. 
Consequently, equations~\eqref{eqn:recursive-3items} and \eqref{eqn:v-moment} enable us to compute Equation~\eqref{eqn:comb-moment} for any given $m$, hence the central moments of the variable $y_{n-1,t}$ of any desired order through a recursive procedure. The procedure begins with the simplest combinations of $(m_1,m_2,m_3)$ where $m=1$, and progresses sequentially to more complex combinations, such as ${(m_1,m_2,m_3), m=2}$, and continues accordingly, adhering to the condition that $m_1+m_2+m_3=m$. Extensions of the recursive approach to more complex models are detailed in \citep{wu2025density}.

While the computational process is conceptually straightforward, it becomes computationally demanding and practically unfeasible for high-order moments if done manually. This computational challenge serves as the impetus for the development of the \pkg{ajdmom} software package, which is designed to automate and streamline the derivation process.

\section[Code design and framework of ajdmom]{Code design and framework of \pkg{ajdmom}}
\subsection{Code design}
Our discussion in the previous section has illustrated the importance of the conditional product moment $\E[I\!E_{n-1,t}^{m_1}I_{n-1,t}^{m_2}I_{n-1,t}^{*m_3}|v_{n-1}]$ in our recursive procedure to calculate the moments. We discover that for any non-negative integer pair $(m_1,m_2,m_3)$, the conditional product moment can be expressed as a generalized polynomial, in the following form:
\begin{align}
    \E[I\!E_{n-1,t}^{m_1}I_{n-1,t}^{m_2}I_{n-1,t}^{*m_3}|v_{n-1}]
    = \sum_{m_1,i,j,l,o,p,q} b_{m_1ijlopq}\cdot x_{m_1ijlopq}\label{eqn:poly}
\end{align}
where $i,j,l,o,p,q$ are integers, and $b_{m_1ijlopq}$ denotes the coefficient associated with the monomial $x_{m_1ijlopq}$ which is defined as
$$
x_{m_1ijlopq} \equiv e^{m_1k(n-1)h} e^{ik[t-(n-1)h]} [t-(n-1)h]^jv_{n-1}^l k^{-o}\theta^p\sigma_v^q.
$$

To manage the complexity of this polynomial form and enable efficient manipulations, we develop a new \proglang{Python} class named \class{Poly}. This class is a specialized dictionary data structure, which extends the capabilities of the \class{UserDict} class in the \proglang{Python} standard library's \pkg{collections} module. The \class{Poly} class is designed specifically to handle the unique requirements of polynomial expressions in our context.

In the context of the polynomial representation given by Equation~\eqref{eqn:poly}, the key computation step within the recursive formula in Equation~\eqref{eqn:recursive-3items} involves the integral:
$$
    \int_{(n-1)h}^t e^{ik[s-(n-1)h]} [s-(n-1)h]^j ds.
$$
For the indefinite integral, the solution can be expressed as:
\begin{equation}\label{eqn:int_et}
\int e^{nkt} t^m dt 
=
\begin{cases}
\sum_{i=0}^m  \frac{c_{nmi}}{k^{i+1}}e^{nkt} t^{m-i}
 & \mathrm{if } n\neq 0, m \neq 0,\\
\frac{1}{nk}e^{nkt}t^0 & \mathrm{if } n\neq 0, m = 0,\\
\frac{1}{m+1}e^{0kt}t^{m+1} & \mathrm{if } n = 0, m \neq 0,\\
e^{0kt}t^1 & \mathrm{if } n =0 , m=0,
\end{cases}
\end{equation}
where $c_{nm0} = 1/n$ and for $1\le i \le m$: 
$$
    c_{nmi} = \frac{(-1)^{i}}{n^{i+1}} \prod_{j=m-i+1}^{m} j.
$$
Thus, the definite integral can be succinctly described by a polynomial of the form:
$$
 \int_{(n-1)h}^t e^{ik[s-(n-1)h]} [s-(n-1)h]^j ds
  = \sum_{i,j',l}b_{ij'l}e^{ik[t-(n-1)h]}[t-(n-1)h]^{j'}k^{-l},
$$
which is represented by an instance of the \class{Poly} class. This instance is essentially a customized dictionary with keys and values corresponding to the polynomial exponents $(i,j',l)$ and coefficients $b_{ij'l}$, and an additional attribute 
\code{keyfor} as the following: 
\begin{Code}
>>> keyfor = ('e^{k[t-(n-1)h]}', '[t-(n-1)h]', 'k^{-}')
\end{Code}
to capture the structure of the polynomial terms.

This approach is leveraged to compute the integral of the expectation, 
$$
    \int_{(n-1)h}^t e^{mks}\E[I\!E_{n-1,s}^{m_1}I_{n-1,s}^{m_2}I_{n-1,s}^{*m_3}|v_{n-1}]ds,
$$
which is needed multiple times in the recursive Equation~\eqref{eqn:recursive-3items}.
The result of this computation is encapsulated by another \class{Poly} object. Accordingly, we have implemented a function named \fct{recursive\_IEII} in the \pkg{ajdmom} package, which executes the recursive process as specified by Equation \eqref{eqn:recursive-3items}.

For the covariance calculations within the package, we focus on the covariance between $y_n$ and its lag-1 counterpart $y_{n+1}$ at different orders, denoted by $\cov(y_n^{l_1}, y_{n+1}^{l_2})$. This necessitates computing $\E[y_n^{l_1}y_{n+1}^{l_2}] - \E[y_n^{l_1}]\E[y_{n+1}^{l_2}]$, which we refer to as having an order of $(l_1,l_2)$. Detailed steps for deriving $\E[y_n^{l_1}y_{n+1}^{l_2}]$ are provided in the \pkg{ajdmom} package documentation, which involves expanding the terms of $y_{n+1}^{l_2}$ followed by those of $y_n^{l_1}$.

The \class{Poly} class, with its custom dictionary structure, has proven crucial for efficiently deriving the moments and covariances for the Heston SV model and its AJD extensions.

\subsection[Framework of the ajdmom package]{Framework of the \pkg{ajdmom} package}
The \pkg{ajdmom} package is methodically organized into two primary components: \textbf{facilities} and \textbf{applications}. Below is an enhanced description of each segment:

\textbf{Facilities}:
The facilities encompass a suite of modules, each dedicated to specific mathematical operations within the domain of stochastic processes:
\begin{itemize}
    \item \emph{poly}: This module introduces the \class{Poly} class, a specialized dictionary-like data structure crafted to represent and manipulate polynomial moments effectively.
    \item \emph{ito\_mom}: This module is dedicated to calculating moments of Itô processes characterized by a single SRD.
    \item \emph{itos\_mom}: An extension of the previous module, \emph{itos\_mom} handles the computation of moments for Itô processes that are derived from a composite of two SRDs.
    \item \emph{ito\_cond\_mom}: This module concentrates on deriving conditional moments of Itô processes from models including jumps in the variance.
    \item \emph{cpp\_mom}: The focus of this module lies in the derivation of moments for compound Poisson processes (CPP).
    \item \emph{utils}: This module provides some support for computations involving combinations and normal distributions.
\end{itemize}

\textbf{Applications}: The applications are divided into subpackages, each targeting a different type of AJD model:
\begin{itemize}
    \item \pkg{mdl\_1fsv}: This subpackage provides tools for the one-factor SV model, notably the Heston SV model.
    \item \pkg{mdl\_1fsvj}: Here, users can find functionality for one-factor SV models that incorporate jumps in the return process.
    \item \pkg{mdl\_2fsv}: Dedicated to two-factor SV models, this subpackage extends the complexity of the SV models.
    \item \pkg{mdl\_2fsvj}: This subpackage is similar to \pkg{mdl\_2fsv} but includes jump features in the return process within the two-factor SV model framework.
    \item \pkg{mdl\_srjd}: This subpackage caters to the square-root jump diffusion (SRJD) process which integrates jump components into the SRD processes.
    \item \pkg{mdl\_svvj}: Extends the Heston SV model by including jumps in the variance process.
    \item \pkg{mdl\_svij}: Enhances the Heston SV model by incorporating independent jumps both in the return and variance processes.
    \item \pkg{mdl\_svcj}: Augments the Heston SV model by adding contemporaneous jumps in the return and variance processes.
\end{itemize}
The \class{Poly} class within the \emph{poly} module is carefully designed to facilitate the representation and manipulation of polynomial moments. It supports arithmetic operations like addition, subtraction, and multiplication via the implementation of corresponding magic methods (a way to overload the behaviors of predefined operators in \proglang{Python}). Additionally, the class allows for a \class{Poly} object to be inversely multiplied by a constant and exponentiated to an integer power, also through magic methods. The documentation of the \pkg{ajdmom} package 
provides an in-depth exploration of these capabilities.

In the \emph{ito\_mom} module, we implement the derivation of the moment $$\E[I\!E_{n-1,t}^{m_1}I_{n-1,t}^{m_2}I_{n-1,t}^{m_3}|v_{n-1}],$$
catering to scenarios where the latent state is represented by an SRD. Similarly, for the case where the latent state is a superposition of two SRDs, the moment 
$$\E[I\!E_{1,n-1,t}^{m_1}I_{1,n-1,t}^{m_2}I\!E_{2,n-1,t}^{m_3}I_{2,n-1,t}^{m_4}I_{n-1,t}^{*m_5}|v_{1,n-1},v_{2,n-1}]$$
is derived within the \emph{itos\_mom} module. The \pkg{ajdmom} package documentation provides detailed definitions of these terms.

The \emph{cpp\_mom} module encompasses the derivation of CPP moments. The default configuration assumes the CPP's arrival process as a homogeneous Poisson process, with the jump distribution typically set to a normal distribution---however, alternative distributions can be accommodated as required. An extension to handle an inhomogeneous Poisson process with a state-dependent affine rate is not covered in the current scope of the package.

Leveraging these facilities, the package simplifies the coding process for automating the derivation of moments and covariances for various AJD models. We have implemented this for four distinct AJD models across four corresponding subpackages. Specifically, the \pkg{mdl\_1fsv} subpackage caters to the one-factor SV model, i.e., the Heston SV model, and is characterized by Equations \eqref{eqn:log-price} and \eqref{eqn:v-t}. Included within this subpackage are three modules---\emph{mom}, \emph{cmom}, and \emph{cov}---for deriving moments, central moments, and covariances, respectively. An additional module, \emph{euler}, facilitates the generation of model samples via the Euler approximation, serving as a validation tool by enabling comparison between derived moments and sample moments. The other three subpackages---\pkg{mdl\_1fsvj}, \pkg{mdl\_2fsv}, and \pkg{mdl\_2fsvj}---mirror this structure, each consisting of modules for moments, central moments, covariances, and sample generation through Euler approximation.

The \pkg{mdl\_1fsvj} subpackage caters to the one-factor SV model with jumps incorporated into returns, governed by the following SDEs:
\begin{align*}
    dp(t)
    &= (\mu - v(t)/2)dt + \sqrt{v(t)}dw^s(t) + dz(t),\\
    dv(t)
    &= k(\theta-v(t))dt + \sigma\sqrt{v(t)}dw^v(t),
\end{align*}
where $z(t)$ denotes a homogeneous CPP with constant arrival rate $\lambda$ and jumps distributed according to $F_j(\cdot,\mb{\theta}_j)$ parameterized by $\mb{\theta}_j$. The remaining parameters are consistent with those in the one-factor SV model as described by Equations~\eqref{eqn:log-price} and \eqref{eqn:v-t}. Within the \pkg{mdl\_1fsvj} subpackage, a normal distribution with mean $\mu_j$ and variance $\sigma_j^2$ serves as a specific instance for the jump size distribution $F_j(\cdot,\mb{\theta}_j)$. 

The \pkg{mdl\_2fsv} subpackage addresses the two-factor SV model which is defined by the following SDEs:
\begin{align*}
    dp(t)
    &= (\mu - v(t)/2)dt + \sqrt{v(t)}dw(t),\\
    v(t) &= v_1(t) + v_2(t),\\
    dv_1(t)
    &= k_1(\theta_1-v_1(t))dt + \sigma_{v1}\sqrt{v_1(t)}dw_1(t),\\
    dv_2(t)
    &= k_2(\theta_2-v_2(t))dt + \sigma_{v2}\sqrt{v_2(t)}dw_2(t),
\end{align*}
where variances $v_1(t)$ and $v_2(t)$ are two independent SRDs, and Wiener processes $w(t)$, $w_1(t)$ and $w_2(t)$ are mutually independent. Extensions to three-factor and higher-factor models are also feasible.

The \pkg{mdl\_2fsvj} subpackage extends the two-factor SV model by incorporating jumps into the return process, articulated by the following SDEs:
\begin{align*}
    dp(t)
    &= (\mu - v(t)/2)dt + \sqrt{v(t)}dw(t) + dz(t),\\
    v(t) &= v_1(t) + v_2(t),\\
    dv_1(t)
    &= k_1(\theta_1-v_1(t))dt + \sigma_{v1}\sqrt{v_1(t)}dw_1(t),\\
    dv_2(t)
    &= k_2(\theta_2-v_2(t))dt + \sigma_{v2}\sqrt{v_2(t)}dw_2(t),
\end{align*}
where $z(t)$ is a CPP analogously defined as in the \pkg{mdl\_1fsvj} subpackage, and all others are set as in the \pkg{mdl\_2fsv} subpackage.

For square-root jump diffusion processes and affine SV models including jumps in the variance \textemdash SV with jumps in the variance (SVVJ), SV with independent jumps in the return and variance (SVIJ), and SV with contemporaneous jumps in the return and variance (SVCJ) \textemdash their moment derivations are much more complicated, see \citet{wu2025density} for detailed explanations. The derivation of unconditional and conditional moments for the SRJD and SVCJ models is implemented in the modules \emph{mom}, \emph{cmom}, \emph{cond\_mom}, \emph{cond\_cmom}, \emph{cond2\_mom} and \emph{cond2\_cmom} within subpackages \pkg{mdl\_srjd} and \pkg{mdl\_svcj}, respectively, where \emph{cond} denotes that only the initial variance is given while \emph{cond2} denotes both the initial variance and jumps over the horizon are given. The derivation of conditional moments and conditional central moments is implemented in the modules \emph{cond2\_mom} and \emph{cond2\_cmom} within subpackages \pkg{mdl\_svvj} and \pkg{mdl\_svij}, respectively. See the package documentation for the details. It should be noted that the derivation of unconditional moments and unconditional central moments for the SVVJ and SVIJ models can also be implemented with moderate adjustments.


The \pkg{mdl\_srjd} subpackage investigates the SRJD process, represented by the following SDE:
\begin{equation*}
    dv(t) = k(\theta - v(t))dt + \sigma_v\sqrt{v(t)}dw^v(t) + dz^v(t),
\end{equation*}
where $z^v(t)$ is a CPP with constant arrival rate $\lambda$ and exponentially distributed jumps with scale parameter $\mu_v$. 

The \pkg{mdl\_svvj} subpackage alters the Heston SV model by substituting its SRD variance with above SRJD process, elucidated by the following SDEs:
\begin{align*}
    dp(t) &= (\mu- v(t)/2) dt + \sqrt{v(t)}dw^s(t),\\
    dv(t) &= k(\theta - v(t))dt + \sigma_v \sqrt{v(t)}dw^v(t) + dz^v(t).
\end{align*}

The \pkg{mdl\_svij} subpackage extends the Heston SV model by incorporating 
independent jumps into the return and variance processes, described by the following SDEs:
\begin{align*}
   dp(t) &= (\mu- v(t)/2) dt + \sqrt{v(t)}dw^s(t) + dz^s(t),\\
   dv(t) &= k(\theta - v(t))dt + \sigma_v \sqrt{v(t)}dw^v(t) + dz^v(t),
\end{align*}
where $z^v(t)$ is a CPP with constant arrival rate $\lambda_v$ and jumps distributed according to an exponential distribution with scale parameter $\mu_v$, $z^s(t)$ is another CPP independent of $z^v(t)$, with constant arrival rate $\lambda_s$ and jumps distributed according to a normal distribution with mean $\mu_s$ and variance $\sigma_s^2$.

The \pkg{mdl\_svcj} subpackage augments the Heston SV model by incorporating 
contemporaneous jumps into the return and variance processes, governed by the following SDEs:
\begin{align*}
   dp(t) &= (\mu- v(t)/2) dt + \sqrt{v(t)}dw^s(t) + dz^s(t),\\
   dv(t) &= k(\theta - v(t))dt + \sigma_v \sqrt{v(t)}dw^v(t) + dz^v(t),
\end{align*}
where $z^s(t)$ and $z^v(t)$ now are two CPPs sharing a common arrival process with constant
arrival rate $\lambda$. Jumps $J_i^v$ in the former CPP are distributed 
according to an exponential distribution with scale parameter $\mu_v$, i.e., 
$J_i^v \sim \mathrm{exp}(\mu_v)$ 
while jumps $J_i^s$ in the latter CPP are
distributed according to a normal distribution with mean $\mu_s + \rho_J J_i^v$ 
(depending on the realized jump $J_i^v$ in the variance process) and variance
$\sigma_s^2$, i.e., $J_i^s \sim \mathcal{N}(\mu_s+\rho_J J_i^v, \sigma_s^2)$.

We note that the supported AJD models within these subpackages are adaptable. For example, the jump size distribution in the subpackage \pkg{mdl\_1fsvj} can be readily generalized to incorporate alternative distributions. While other extensions are conceivable, they may require additional exploration and validation.

\section{Usage demonstrations and experiments}
In this section, we first use the Heston SV model as an illustrative example to demonstrate the application of the \pkg{ajdmom} package. Second, we present a series of numerical experiments to validate the accuracy and reliability of the methodologies we propose.
Before using the \pkg{ajdmom} package, we first need to install it from \proglang{Python} package index (PyPI),  through
\begin{Code}
> pip install ajdmom
\end{Code}
or some other ways.

\subsection{Usage demonstrations}
As a case study, we utilize the Heston SV model, a representative of the one-factor SV model category, to elucidate the functionality of the \pkg{ajdmom} package. In the literature, the Heston SV model, as rigorously defined and adopted in this research, has limited closed-form solutions for its moments and covariances, with few exceptions such as the expected value $\E[y_n] = (\mu-\theta/2)h$. We aim to demonstrate that the \pkg{ajdmom} package is capable not only of replicating these known results but also of facilitating the computation of hitherto unattainable outcomes.

To illustrate this, we begin by detailing the process for calculating the first moment of the Heston SV model. The procedure is exemplified through the following code excerpt:
\begin{CodeChunk}
\begin{CodeInput}
>>> from ajdmom import mdl_1fsv
>>> from pprint import pprint
>>> m1 = mdl_1fsv.moment_y(1)
>>> msg = f"which is a Poly with attribute keyfor = \n{m1.keyfor}"
>>> print("moment_y(1) = "); pprint(m1); print(msg)
\end{CodeInput}
\begin{CodeOutput}
moment_y(1) = 
{(0, 1, 0, 0, 1, 0, 0, 0): Fraction(-1, 2),
 (0, 1, 0, 1, 0, 0, 0, 0): Fraction(1, 1)}
which is a Poly with attribute keyfor = 
('e^{-kh}', 'h', 'k^{-}', 'mu', 'theta', 'sigma_v', 'rho', 'sqrt(1-rho^2)')
\end{CodeOutput}
\end{CodeChunk}
The code snippet's execution yields a \class{Poly} object which contains two principal key-value pairs, namely (0,1,0,0,1,0,0,0): Fraction(-1,2) and (0,1,0,1,0,0,0,0): Fraction(1,1), correspond to the following expressions:
\begin{align*}
    -\frac{1}{2}\times & e^{-0kh}h^1k^{-0}\mu^0\theta^1\sigma_v^0\rho^0\left(\sqrt{1-\rho^2}\right)^0,\\
    1\times & e^{-0kh}h^1k^{-0}\mu^1\theta^0\sigma_v^0\rho^0\left(\sqrt{1-\rho^2}\right)^0,
\end{align*}
respectively. The summation of these terms yields the first moment of the one-factor SV model: $\E[y_n] = (\mu-\theta/2)h$. This demonstrates that the \pkg{ajdmom} package successfully encapsulates the model's dynamics into a computationally manipulable form, specifically leveraging a custom dictionary data structure, referred to as \class{Poly}, to encode the moment's expression.

Subsequently, we shift our focus to the computation of the covariance $\cov(y_n^2,y_{n+1})$ for the Heston SV model---an outcome not previously resolved in the literature. This computation is facilitated by the following code snippet: 
\begin{CodeChunk}
\begin{CodeInput}
>>> from ajdmom import mdl_1fsv 
>>> from pprint import pprint
>>> cov21 = mdl_1fsv.cov_yy(2,1)
>>> msg = f"which is a Poly with attribute keyfor =\n{cov21.keyfor}"
>>> print("cov_yy(2,1) = "); pprint(cov21); print(msg)
\end{CodeInput}
\begin{CodeOutput}
cov_yy(2,1) =
{(0, 0, 3, 0, 1, 2, 0, 2): Fraction(-1, 4),
 (0, 0, 3, 0, 1, 2, 2, 0): Fraction(-5, 4),
 (0, 0, 4, 0, 1, 3, 1, 0): Fraction(3, 4),
 (0, 0, 5, 0, 1, 4, 0, 0): Fraction(-1, 8),
 (0, 1, 2, 0, 2, 1, 1, 0): Fraction(1, 2),
 (0, 1, 2, 1, 1, 1, 1, 0): Fraction(-1, 1),
 (0, 1, 3, 0, 2, 2, 0, 0): Fraction(-1, 8),
 (0, 1, 3, 1, 1, 2, 0, 0): Fraction(1, 4),
 (1, 0, 3, 0, 1, 2, 0, 2): Fraction(1, 2),
 (1, 0, 3, 0, 1, 2, 2, 0): Fraction(5, 2),
 (1, 0, 4, 0, 1, 3, 1, 0): Fraction(-3, 2),
 (1, 0, 5, 0, 1, 4, 0, 0): Fraction(1, 4),
 (1, 1, 2, 0, 1, 2, 2, 0): Fraction(1, 1),
 (1, 1, 2, 0, 2, 1, 1, 0): Fraction(-1, 1),
 (1, 1, 2, 1, 1, 1, 1, 0): Fraction(2, 1),
 (1, 1, 3, 0, 1, 3, 1, 0): Fraction(-3, 4),
 (1, 1, 3, 0, 2, 2, 0, 0): Fraction(1, 4),
 (1, 1, 3, 1, 1, 2, 0, 0): Fraction(-1, 2),
 (1, 1, 4, 0, 1, 4, 0, 0): Fraction(1, 8),
 (2, 0, 3, 0, 1, 2, 0, 2): Fraction(-1, 4),
 (2, 0, 3, 0, 1, 2, 2, 0): Fraction(-5, 4),
 (2, 0, 4, 0, 1, 3, 1, 0): Fraction(3, 4),
 (2, 0, 5, 0, 1, 4, 0, 0): Fraction(-1, 8),
 (2, 1, 2, 0, 1, 2, 2, 0): Fraction(-1, 1),
 (2, 1, 2, 0, 2, 1, 1, 0): Fraction(1, 2),
 (2, 1, 2, 1, 1, 1, 1, 0): Fraction(-1, 1),
 (2, 1, 3, 0, 1, 3, 1, 0): Fraction(3, 4),
 (2, 1, 3, 0, 2, 2, 0, 0): Fraction(-1, 8),
 (2, 1, 3, 1, 1, 2, 0, 0): Fraction(1, 4),
 (2, 1, 4, 0, 1, 4, 0, 0): Fraction(-1, 8)}
which is a Poly with attribute keyfor =
('e^{-kh}', 'h', 'k^{-}', 'mu', 'theta', 'sigma_v', 'rho', 'sqrt(1-rho^2)')
\end{CodeOutput}
\end{CodeChunk}
The execution of the provided code yields the covariance expressed within a \class{Poly} object.

The \pkg{ajdmom} package further extends its utility to encompass seven additional extended models, which are accessible through the subpackages \pkg{mdl\_1fsvj}, \pkg{mdl\_2fsv}, \pkg{mdl\_2fsvj}, \pkg{mdl\_srjd}, \pkg{mdl\_svvj}, \pkg{mdl\_svij} and \pkg{mdl\_svcj}, respectively. The representation of moments and covariances as \class{Poly} objects simplifies the process of performing partial differentiation with respect to the model parameters. Detailed instructions for these operations are available in the \pkg{ajdmom} package documentation.

\subsection{Experiments}
To evaluate the accuracy of our proposed methods and their implementation, we performed a series of experiments under the SVJ and SVCJ models, as implemented in the \pkg{mdl\_1fsvj} and \pkg{mdl\_svcj} subpackages, respectively.

The initial experimental set aimed to validate the moment calulations provided by the \pkg{ajdmom} package. This was achieved by comparing the theoretically derived moments against their sample counterparts, computed from a sample path of 4,000,000 observations of the model under the parameterization $\mu = 0.125$, $k_1 = 0.1$, $\theta_1 =0.25$, $\sigma_{v1}=0.1$, $\lambda=0.01$, $\mu_j=0$, $\sigma_j=0.05$. The sample path was generated using Euler's method, with a step size $h=1$, further subdivided into ten equal segments for finer approximation. Table~\ref{tab:comp-mom-1fsvj} represents the comparative results, affirming the correctness of our method and implementation within the expected variability due to sampling randomness and the discretization error inherent in Euler's approximation.

\begin{table}[t!]
  \centering
  \begin{tabular}{ccccc}
    \hline
    Moment& Derived moment & Sample moment & Difference & Difference (\%)\\
    \hline
    $\E[y_n]$ &~0.0000 &~0.0002 &0.0002 &-\\
    $\E[y_n^2]$ &~0.2615 &~0.2606 &0.0009 &0\%\\
    $\E[y_n^3]$ &-0.0449 &-0.0425 &0.0024 &5\%\\
    $\E[y_n^4]$ &~0.2508 &~0.2491 &0.0016 &1\%\\
    $\E[y_n^5]$ &-0.1412 &-0.1350 &0.0062 &4\%\\
    \hline
  \end{tabular}
    \caption{\label{tab:comp-mom-1fsvj} Comparison between derived moments and sample moments for the one-factor SV model with jumps in the return process.  The comparison is based on a sample path of 4,000,000 observations of the one-factor SV model with jumps in the return process, under the parameter settings ($\mu = 0.125$, $k_1 = 0.1$, $\theta_1 =0.25$, $\sigma_{v1}=0.1$, $\lambda=0.01$, $\mu_j=0$, $\sigma_j=0.05$). The sample path was generated using Euler's method, with a step size $h=1$, further subdivided into ten equal segments for finer approximation.}
\end{table}

The subsequent experimental series examined the accuracy of the covariance calculations. Keeping all settings consistent with the first series, we shifted our focus to the derivation of covariances. The outcomes, as summarized in Table~\ref{tab:comp-cov-1fsvj}, likewise support the validity of our methods and their computational realizations. For the SVCJ model, we conduct experiments to test the accuracy of the conditional moment computation, with results reported in Table~\ref{tab:comp-mom-svcj}.

\begin{table}[t!]
    \centering
    \begin{tabular}{ccccc}
         \hline
         Covariance& Derived Covariance & Sample Covariance & Difference & Difference (\%)\\
         \hline
         $\cov(y_n,y_{n+1})$ &~0.0108 &~0.0108 &0.0000 &0\%\\
         $\cov(y_n^2,y_{n+1})$ &-0.0069 &-0.0069 &0.0000 &0\%\\
         $\cov(y_n,y_{n+1}^2)$ &-0.0228 &-0.0230 &0.0002 &1\%\\
         $\cov(y_n^3,y_{n+1})$ &~0.0112 &~0.0111 &0.0001 &1\%\\
         $\cov(y_n^2,y_{n+1}^2)$ &~0.0150 &~0.0150 &0.0000 &0\%\\
         $\cov(y_n,y_{n+1}^3)$ &~0.0140 &~0.0138 &0.0002 &1\%\\
         $\cov(y_n^4,y_{n+1})$ &-0.0155 &-0.0151 &0.0004 &2\%\\
         $\cov(y_n^3,y_{n+1}^2)$ &-0.0243 &-0.0245 &0.0002 &1\%\\
         $\cov(y_n^2,y_{n+1}^3)$ &-0.0108 &-0.0108 &0.0000 &0\%\\
         $\cov(y_n,y_{n+1}^4)$ &-0.0456 &-0.0456 &0.0000 &0\%\\
         \hline
  \end{tabular}
  \caption{\label{tab:comp-cov-1fsvj} Comparison between derived covariances and sample covariances for the one-factor SV model with jumps in the return process. The comparison is based on a sample path of 4,000,000 observations of the one-factor SV model with jumps in the return process, under the parameter settings ($\mu = 0.125$, $k_1 = 0.1$, $\theta_1 =0.25$, $\sigma_{v1}=0.1$, $\lambda=0.01$, $\mu_j=0$, $\sigma_j=0.05$). The sample path was generated using Euler's method, with a step size $h=1$, further subdivided into ten equal segments for finer approximation.}
\end{table}

\begin{table}[t!]
  \centering
  \begin{tabular}{ccccc}
    \hline
    Moment& Derived moment & Sample moment & Difference & Difference (\%)\\
    \hline
    $\E[y_n|v_0]$ &~0.0229 &~0.0228 &0.0001 &1\%\\
    $\E[y_n^2|v_0]$ &~0.0196 &~0.0200 &0.0003 &2\%\\
    $\E[y_n^3|v_0]$ &-0.0024 &-0.0025 &0.0001 &5\%\\
    $\E[y_n^4|v_0]$ &~0.0022 &~0.0023 &0.0001 &5\%\\
    $\E[y_n^5|v_0]$ &-0.0011 &-0.0012 &0.0001 &7\%\\
    \hline
  \end{tabular}
    \caption{\label{tab:comp-mom-svcj} Comparison between derived conditional moments and sample moments for the SV model with contemporaneous jumps in the return and the volatility.  The comparison is based on 4,000,000 i.i.d. samples of the SVCJ model, under the parameter settings ($v_0 = 0.007569$, $\mu = 0.0789$, $k=3.46$, $\theta = 0.008$, $\sigma = 0.14$, $\rho = -0.82$, $\lambda = 0.47$, $\mu_v = 0.05$, $\rho_J = -0.38$, $\mu_s = -0.0865$, $\sigma_s = 0.0001$). The samples are generated using Euler's method, with $h=1$, subdivided into ten equal segments for fine approximation.}
\end{table}

All of the tables corroborate the effectiveness of our procedures, demonstrating a high degree of concordance between the theoretical derivations and their empirical estimates. These findings are particularly encouraging considering the stochastic nature of the model and the approximations employed in the simulation process.

\section{Conclusion}
In this work, we have developed and introduced \pkg{ajdmom}, a comprehensive \proglang{Python} package designed for the derivation of moment formulae specifically tailored to the Heston SV model and its AJD extensions.

Our approach commenced with an exposition of the AJD models, highlighting the Heston SV model as a prominent example. We then present a recursive equation framework capable of determining moments of any order within these models. Leveraging insights from this theoretical foundation, we crafted a novel dictionary-based data structure to effectively represent the moments of the It\^{o} processes under consideration.

The paper proceeded to detail the implementation of a procedure within \pkg{ajdmom} to automate the derivation of moments for AJD models. We discussed the architectural framework of the package, emphasizing its robustness and versatility.

Practical applications and the utility of \pkg{ajdmom} were demonstrated through illustrative examples, and experimental validation confirmed the accuracy of our methodology. The results not only verified the correctness of our proposed solutions but also underscored the practical implications for users involved in financial modelling and analysis.

Looking ahead, the methodologies encapsulated within \pkg{ajdmom} hold the promise for extension to a broader class of AJD models, including potentially non-affine jump diffusion processes. Such ventures present intriguing avenues for future research and development in this domain.

In summary, \pkg{ajdmom} stands as a significant contribution to the computational finance field, offering a powerful tool for researchers and practitioners alike, and paving the way for continued innovation in the modelling of complex financial instruments. 

\section*{Acknowledgement}
This work is supported in part by the National Nature Science Foundation of China (NSFC) under grants 72033003, 72350710219 and 72342006.

\bibliography{references}


\newpage

\begin{appendix}
\section{Recursive equation for the product moments}

The conditional product moments of It\^{o} processes $I\!E_{n-1,t}^{m_1}I_{n-1,t}^{m_2}I_{n-1,t}^{*m_3}$, with condition on $v_{n-1}$, can be derived in a recursive way as follows:
\begin{align*}
&\E[I\ie_{n-1,t}^{m_1}I_{n-1,t}^{m_2}I_{n-1,t}^{*m_3}|v_{n-1}]\\
&= \int_{(n-1)h}^t e^{ks} \E[I\!E_{n-1,s}^{m_1-2}I_{n-1,s}^{m_2}I_{n-1,s}^{*m_3}|v_{n-1}]ds \cdot \frac{m_1(m_1-1)}{2}e^{k(n-1)h}\bar{v}_{n-1}\\
&\quad + \int_{(n-1)h}^t e^{2ks} \E[I\!E_{n-1,s}^{m_1-2}I_{n-1,s}^{m_2}I_{n-1,s}^{*m_3}|v_{n-1}]ds \cdot \frac{m_1(m_1-1)}{2}\theta\\
&\quad +  \int_{(n-1)h}^t e^{ks} \E[I\!E_{n-1,s}^{m_1-1}I_{n-1,s}^{m_2}I_{n-1,s}^{*m_3}|v_{n-1}]ds \cdot \frac{m_1(m_1-1)}{2}\sigma_v\\
&\quad + \int_{(n-1)h}^t e^{-ks} \E[I\!E_{n-1,s}^{m_1}I_{n-1,s}^{m_2-2}I_{n-1,s}^{*m_3}|v_{n-1}]ds \cdot \frac{m_2(m_2-1)}{2}e^{k(n-1)h}\bar{v}_{n-1} \\
&\quad + \int_{(n-1)h}^t \E[I\!E_{n-1,s}^{m_1}I_{n-1,s}^{m_2-2}I_{n-1,s}^{*m_3}|v_{n-1}]ds \cdot \frac{m_2(m_2-1)}{2}\theta\\
&\quad + \int_{(n-1)h}^t e^{-ks} \E[I\!E_{n-1,s}^{m_1+1}I_{n-1,s}^{m_2-2}I_{n-1,s}^{*m_3}|v_{n-1}]ds \cdot \frac{m_2(m_2-1)}{2}\sigma_v \\
&\quad + \int_{(n-1)h}^t  \E[I\!E_{n-1,s}^{m_1-1}I_{n-1,s}^{m_2-1}I_{n-1,s}^{*m_3}|v_{n-1}]ds \cdot m_1m_2 e^{k(n-1)h}\bar{v}_{n-1} \\
&\quad + \int_{(n-1)h}^t  e^{ks}\E[I\!E_{n-1,s}^{m_1-1}I_{n-1,s}^{m_2-1}I_{n-1,s}^{*m_3}|v_{n-1}]ds \cdot m_1m_2\theta\\
&\quad + \int_{(n-1)h}^t \E[I\!E_{n-1,s}^{m_1}I_{n-1,s}^{m_2-1}I_{n-1,s}^{*m_3}|v_{n-1}]ds \cdot m_1m_2\sigma_v\\
&\quad + \int_{(n-1)h}^t e^{-ks} \E[I\!E_{n-1,s}^{m_1}I_{n-1,s}^{m_2}I_{n-1,s}^{*m_3-2}|v_{n-1}]ds \cdot m_3 e^{k(n-1)h}\bar{v}_{n-1} \\
&\quad + \int_{(n-1)h}^t \E[I\!E_{n-1,s}^{m_1}I_{n-1,s}^{m_2}I_{n-1,s}^{*m_3-2}|v_{n-1}]ds \cdot m_3\theta\\
&\quad + \int_{(n-1)h}^t e^{-ks} \E[I\!E_{n-1,s}^{m_1+1}I_{n-1,s}^{m_2}I_{n-1,s}^{*m_3-2}|v_{n-1}]ds \cdot m_3\sigma_v.
\end{align*}

\end{appendix}


\end{document}